\documentstyle[epsf]{elsart}

\newcommand{\beq}{\begin{equation}}
\newcommand{\eeq}{\end{equation}}
\newcommand{\bea}{\begin{eqnarray}}
\newcommand{\eea}{\end{eqnarray}}
\newcommand{\beas}{\begin{eqnarray*}}
\newcommand{\eeas}{\end{eqnarray*}}

\renewcommand{\Rset}{{\mbox{$I  \! \! R$}}}

\begin{document}

\begin{frontmatter}
\title{``Embedded solitons": solitary waves in resonance with the linear
spectrum \thanksref{draft}}

\author{A.R.\ Champneys$^{(1)}$, B.A.\ Malomed$^{(2)}$ J.\ Yang$^{(3)}$ and
D.J.\ Kaup$^{(4)}$ }

\address{(1) Department of Engineering Mathematics, University of Bristol,
Bristol BS8DITR, U.K\\
(2) Faculty of Engineering, Tel Aviv University, Tel Aviv 69978, Israel\\
(3) Department of Mathematics and Statistics, University of Vermont, Burlington
VT 05401, USA \\
(4) Clarkson University, Potsdam, NY 13699-5815, USA\\ }

\thanks[draft]{To appear in a special issue of Physica D;
Proceedings of a meeting devoted to the 60th birthday of V.E. Zakharov
(Chernogolovka, Russia, 1999)}

\end{frontmatter}

{\bf Abstract}

It is commonly held that a necessary condition for the existence of solitons
in nonlinear-wave systems is that the soliton's frequency (spatial or
temporal) must not fall into the continuous spectrum of radiation modes.
However, this is not always true. We present a new class of {\it %
codimension-one} solitons (i.e., those existing at isolated frequency
values) that are {\em embedded} into the continuous spectrum. This is
possible if the spectrum of the linearized system has (at least) two
branches, one corresponding to exponentially localized solutions, and the
other to radiation modes. An embedded soliton (ES) is obtained when the
latter component exactly vanishes in the solitary-wave's tail. The paper
contains both a survey of recent results obtained by the authors and some
new results, the aim being to draw together several different mechanism
underlying the existence of ESs. We also consider the distinctive property
of {\em semi-stability} of ES, and {\em moving} ESs. Results are presented
for four different physical models, including an extended 5th-order KdV
equation describing surface waves in inviscid fluids, and three models from
nonlinear optics. One of them pertains to a resonant Bragg grating in an
optical fiber with a cubic nonlinearity, while two others describe
second-harmonic generation (SHG) in the temporal or spatial domain (i.e.,
respectively, propagating pulses in nonlinear optical fibers, or stationary
patterns in nonlinear planar waveguides). Special attention is paid to the
SHG model in the temporal domain for a case of competing quadratic and cubic
nonlinearities. An essential new result is that ES is, virtually, fully
stable in the latter model in the case when both harmonics have anomalous
dispersion.

\section{Introduction}

Recent studies have revealed a novel type of solitary waves (which we will
loosely called ``solitons'', without assuming integrability of the
underlying models) that are {\it embedded} into the continuous spectrum,
i.e., the soliton's internal frequency is in resonance with linear
(radiation) waves. Generally, such a soliton should not exist, one finding
instead a quasi- (delocalized) soliton with nonvanishing oscillatory tails
(radiation component) \cite{Bo:98}. Nevertheless, {\em bona fide}
(exponentially decaying) solitons can exist as {\it codimension-one}
solutions if, at discrete values of the (quasi-)soliton's internal
frequency, the amplitude of the tail exactly vanishes, while the soliton
remains embedded into the continuous spectrum of the radiation modes. This
requires the spectrum of the corresponding linearized system to consist of
(at least) two branches, one corresponding to exponentially localized
solutions, and the other to oscillatory radiation modes. In terms of the
corresponding ordinary differential equations (ODEs) for the traveling-wave
solutions, the origin must be a {\em saddle-centre}, that is its
linearisation gives rise to both real and pure imaginary eigenvalues.

Examples of such {\it embedded solitons} (ESs) were found in water-wave
models \cite{ChGr:96} and in several nonlinear-optical ones, including a
Bragg-grating model with the wave-propagation (second-order-derivative)
linear terms taken into regard \cite{ChMaFr:98}, and a model of the
second-harmonic generation (SHG) in the presence of a Kerr nonlinearity \cite
{YaMaKa:99}. The term ``ES'' was proposed in the latter work.

ESs are interesting for several reasons, firstly because they frequently
appear when higher-order (singular) perturbations are added to the system,
which may completely change its soliton spectrum (see, e.g.,\ \cite
{ChMaFr:98}). Secondly, optical ESs have considerable potential for
applications, just because they are isolated solitons, rather than members
of continuous families. Finally, and most crucial for physical applications,
ESs are {\em semi}-stable objects. That is, as argued in Ref.\ \cite
{YaMaKa:99} analytically, in a general form applicable to ESs in any system,
and checked numerically for the SHG model with the additional defocusing
Kerr nonlinearity, ESs are stable in the linear approximation, but do have a
slowly growing (sub-exponential) {\em one-sided} nonlinear instability (see
Section \ref{sec:3} below).

In the next section we discuss the existence and stability of ESs in four
different nonlinear partial-differential-equation (PDE) models.
Mathematically speaking, in each case the reduced traveling-wave or
steady-state ODEs have the structure of a fourth-order, reversible
Hamiltonian system. In accord with what was said above, a parameter region
we focus on is where the origin (trivial fixed point) in these ODE systems
is a saddle-centre. That is, after diagonalizing the system, one
two-dimensional (2D) component of the dynamical system gives rise to
imaginary eigenvalues $\pm i\omega $ (corresponding to a continuous
radiation branch in the linear spectrum of the PDE system), and the other to
real eigenvalues $\pm \lambda $ (corresponding to a gap in the radiation
spectrum).

All four models considered in this paper share the feature that ESs exist as
non-generic, codimension-one solutions. Section 3 then presents three
different views as to why this should be, which together provide for general
insight into the existence and multiplicity of ESs. Section \ref{sec:3} goes
on to discuss their stability, arguing that ESs, in general, may be
neutrally stable linearly, but suffer from a one-sided sub-exponential
instability. This situation has been termed {\em semi-stability} in 
\cite{YaMaKa:99}. Section \ref{sec:4} treats ``moving" embedded solitons and,
finally, Section (\ref{sec:5}) draws conclusions and briefly discusses
potential physical applications of ESs in optical memory devices.

\section{Physical Examples}

\subsection{An extended 5th-order KdV equation}

One of the first systems in which ESs 
were found (although without being given that
name) is an extended 5th-order Korteweg - de Vries (KdV) equation \cite
{ChGr:96,KiOl:92}, 
\begin{equation}
u_{t}=\left[ (2/15)u_{xxxx}-bu_{xx}+au+(3/2)u^{2}+\mu \left(
(1/2)(u_{x})^{2}+(uu_{x})_{x}\right) \right ]_{x},  \label{5thKdV}
\end{equation}
which with $\mu=0$ reduces to the usual 5th-order KdV equation studied by a
number of authors, see, e.g., Refs. \cite{HuSc:88,Ka:94a}. The extended form
(\ref{5thKdV}) may be derived via a regular Hamiltonian perturbation theory
from an exact Euler-equation formulation for water-waves with surface
tension \cite{CrGr:94}. Looking for traveling-wave solutions $u(x-ct)$,
integrating once, setting the constant of integration to be zero, and
absorbing the linear term $\sim u_{x}$ by redefining $a$, one arrives at the
following ODE (the prime stands for $d/d(x-ct)$), 
\begin{equation}
\frac{2}{15}u^{\prime \prime \prime \prime }-bu^{\prime \prime }+au+\frac{3}{%
2}u^{2}+\mu \left[ \frac{1}{2}(u^{\prime })^{2}+(uu^{\prime })^{\prime }%
\right] =0.  \label{grovesmodel}
\end{equation}

When $a<0$, Eq. (\ref{grovesmodel}) is in the ES regime, since the
linearization around the origin, $u=0$, yields both real and imaginary
eigenvalues. Note that, for the particular case $\mu =0$, it has been proved
that there are no dynamical orbits homoclinic to the origin \cite{AmMcK:91}.
Nevertheless, at least in the limit $a\rightarrow -0$, the system does
possess a large family of homoclinic connections to periodic orbits, rather
than to fixed points \cite{GrJo:95}. The latter generic solutions correspond
to the above-mentioned delocalized solitary waves \cite{Bo:98}.

\begin{figure}[tbp]
\centerline{\epsfxsize 4in
\epsffile{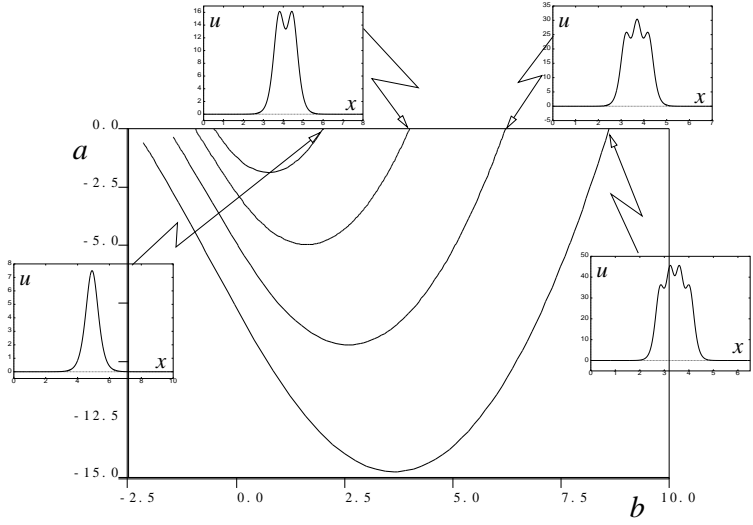}}
\caption{{\protect\small The region of existence of the embedded solitons
for Eq. $(\ref{grovesmodel})$}}
\label{F:groves_gen}
\end{figure}

However, ESs {\em do} exist in the case $\mu =1$ \cite{ChGr:96}, where one
can find the {\em explicit} family of homoclinic-to-zero solutions, 
\begin{equation}
u(t)=3\left( b+\frac{1}{2}\right) \mbox{sech}^{2}\left( \sqrt{\frac{3(2b+1)}{%
4}}t\right), \: a=\frac{3}{5}(2b+1)(b-2), \: b\geq -1/2.  \label{KdVexplicit}
\end{equation}
Moreover, this curve (family) of ESs was found in \cite{ChGr:96} to be only
the first in a countable set of curves that appear to bifurcate from $a=0$
at a discrete set of negative values of $b$, see Fig. \ref{F:groves_gen}.
Note that there are numerical difficulties in computing up to the limit
point $a=0$ for $b<0$; this is because, as will be motivated in Section \ref
{sec:2.3} below, the bifurcation of these solutions from $a=0$ is a
`beyond-all-orders' effect

We remark that Grimshaw and Cook \cite{GrCo:96} found a similar
bifurcation-type phenomenon in a system of two coupled KdV equations,
although they did not explicitly identify the vanishing of the tail
amplitude of the delocalized solitary waves as what we now call ESs. Also,
Fujioka and Espinosa \cite{FuEs:96} found a single {\em explicit} ES in a
higher-order NLS equation with a quintic nonlinearity.

Note also the following feature of this family of ES states. The first
member of the family, viz., the explicit solution (\ref{KdVexplicit})) is a
fundamental ``ground state", while all other solution branches represent `
``excited states", having the form of the ground-state soliton with
small-amplitude ``ripples" superimposed on it. As yet unpublished numerical
results suggest that the ground-state soliton appears to be dynamically
stable (in the sense described in Section \ref{sec:3} below).

\subsection{A generalized Massive Thirring model}

The first example of ESs in nonlinear optics was found in a generalized
Thirring model (GTM), introduced long ago in \cite{ChJo:89,AcWa:89}. ESs
appear in this model when additional wave-propagation (second-derivative)
terms are included ($k$ being the carrier wavenumber), so that the model
takes the form 
\begin{equation}
\begin{array}{rcl}
iu_{t}+iu_{x}+(2k)^{-1}\left( u_{xx}-u_{tt}\right) +\left[
(1/2)|u|^{2}+|v|^{2}\right] u+v & = & 0, \\ 
iv_{t}-iv_{x}+(2k)^{-1}\left( v_{xx}-v_{tt}\right) +\left[
(1/2)|v|^{2}+|u|^{2}\right] v+u & = & 0.
\end{array}
\label{GTMPDE}
\end{equation}
Here $u(x,t)$ and $v(x,t)$ are right- and left- traveling waves coupled by
resonant reflections on the grating. Soliton solutions are sought for as $%
u(x,t)=\exp (-i\Delta \omega t)\,U(\xi )$, $\;v(x,t)=\exp (-i\Delta \omega
t)\,V(\xi )$, where $\xi \equiv x-ct$, $c$ and $\Delta \omega $ being
velocity and frequency shifts. The functions $U(\xi )$ and $V(\xi )$ satisfy
ODEs 
\begin{equation}
\begin{array}{rcl}
\chi U+i(1-C)U^{\prime }+DU^{\prime \prime }+\left[ (1/2)|U|^{2}+|V|^{2} 
\right] U +V & = & 0, \\ 
\chi V-i(1+C)V^{\prime }+DV^{\prime \prime }+\left[ (1/2)|V|^{2}+|U|^{2} 
\right] V+U & = & 0,
\end{array}
\label{GTM}
\end{equation}
where $\chi \equiv \Delta \omega +\left( \Delta \omega \right) ^{2}/2k$, the
effective velocity is $C\equiv (1+\Delta \omega /k)c$, and an effective {\it %
dispersion coefficient} is $D\equiv \left( 1-c^{2}\right) /2k$. The same
equations were derived in \cite{ChMaFr:98} for: (i) temporal-soliton
propagation in nonlinear fiber gratings, including spatial-dispersion
effects; and (ii) spatial solitons in a planar waveguide with a Bragg
grating in the form of parallel scores, taking diffraction into regard, with 
$t$ replaced by the propagation coordinate $z$, while $x$ is the transverse
coordinate. If we set $C=0$ then Eqs. (\ref{GTM}) admit the further
invariant reduction $U=V^{\ast }$, leading to a single ODE \cite{ChMaFr:98}, 
\begin{equation}
DU^{\prime \prime }+iU^{\prime }+\chi U+(3/2)|U|^{2}U+U^{\ast }=0,
\label{GTM4thODE}
\end{equation}
which is equivalent to a (Hamiltonian and reversible) system of four
first-order equations for the real and imaginary parts of $U$ and $U^{\prime
}$. In this system, the origin is a saddle-centre, provided that $D>0$ and $%
|\chi |<1$. In \cite{ChMaFr:98} it was found numerically that Eq. (\ref
{GTM4thODE}) admits exactly {\em three} branches of the fundamental ESs (in
contrast to Eq. (\ref{5thKdV}) where countably many ES states have been
found). Also {\it moving} ($c\neq 0$) ESs, satisfying the 8th-order system $(%
\ref{GTM})$, have also recently been found in Ref. \cite{ChMa:99a}; see
Section \ref{sec:4} below for more details.

\subsection{A three-wave interaction model}

\begin{figure}[tbp]
\epsfxsize 5.5in \centerline{\epsffile{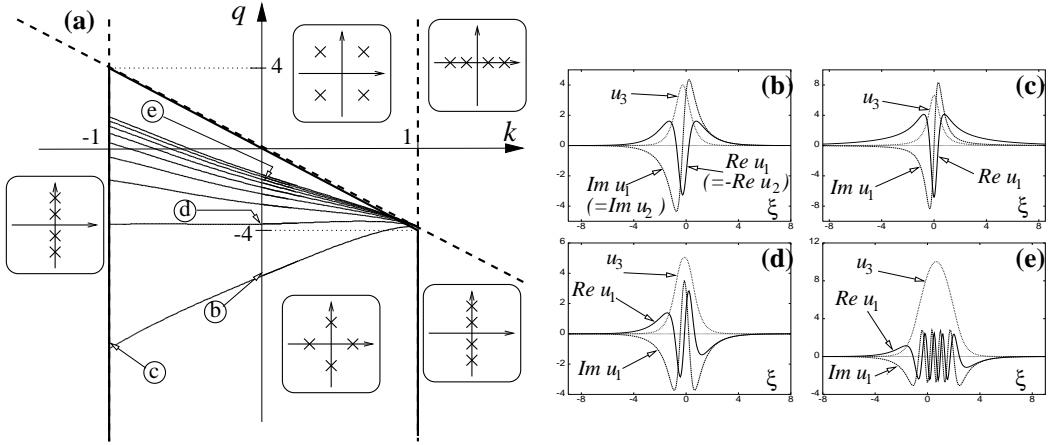}}
\caption{{\protect\small (a) The $(k,q)$ parameter plane of the three-wave
model $(\ref{MakPDE})$. The linear analysis (summarized in the inset boxes)
shows that ESs can occur only in the region between the bold lines. The
bundle of curves emanating from the point $(k=1,q=-4)$ are branches of
embedded-soliton solutions with $c=0$. The panels (b)-(e) depict solutions
at the labeled points.}}
\label{F:Mak0}
\end{figure}

ESs can be found in far greater abundance in a model for {\em spatial
solitons}, assuming a planar waveguide with a quadratic ($\chi ^{(2)}$)
nonlinearity \cite{MaMaCh:98}, where two {\em fundamental-harmonic} (FH)
waves $v_{1,2}$ are coupled by the Bragg reflections from a set of parallel
scores. These two waves then interact nonlinearly, and generate a third
wave, the {\it second-harmonic} (SH), with its wave-vector equal to the sum
of those of the two FH components. The set of equations are: 
\begin{equation}
\begin{array}{rcl}
i(v_{1,2})_{z}\pm i(v_{1,2})_{x}+v_{2,1}+v_{3}v_{2,1}^{\ast } & = & 0, \\ 
2i(v_{3})_{z}-qv_{3}+D(v_{3})_{xx}+v_{1}v_{2} & = & 0.
\end{array}
\label{MakPDE}
\end{equation}
Here $v_{3}$ is the SH field, $x$ is the normalized transverse coordinate, $q
$ is a mismatch parameter, and $D$ is an effective diffraction coefficient.
Solutions to Eq.\ (\ref{MakPDE}) are sought in the form $v_{1,2}(x,z)=\exp
(ikz)\,u_{1,2}(\xi )$, $\;v_{3}(x,z)=\exp (2ikz)\,u_{3}$, with $\xi \equiv
x-cz$, $c$ being the slope of the soliton's axis relative to the propagation
direction $z$. In Ref. \cite{ChMa:99b}, many ESs of the zero-walkoff type ($%
c=0$) were found (summarized in Fig.\ \ref{F:Mak0}), in which case the ODEs
reduce, as in Eqs. (\ref{GTM}), to a fourth-order real system. When $c\neq 0$%
, one cannot assume all the amplitudes to be real. In this case, one finds
``moving" ESs as solutions to an eighth-order real ODE system (see Section 
\ref{sec:4} below).

\subsection{A second-harmonic-generation system}

The model we shall study in most detail is that for which the term `ES' was
first proposed in \cite{YaMaKa:99}, viz., a nonlinear optical medium with
competing quadratic and cubic nonlinearities \cite{TrBuKi:96,BaKiBu:97} 
\begin{equation}
\begin{array}{rcl}
iu_{z}+(1/2)u_{tt}+u^\ast v+\gamma _{1}(|u|^{2}+2|v|^{2})u & = & 0, \\ 
iv_{z}-(1/2)\delta v_{tt}+qv+(1/2)u^{2}+2\gamma _{2}(|v|^{2}+2|u|^{2})v & =
& 0.
\end{array}
\label{SHGPDE}
\end{equation}
Here, $u$ and $v$ are FH and SH amplitudes, $-\delta $ is a relative SH/FH
dispersion coefficient, $q$ is a phase-velocity mismatch, and $\gamma _{1,2}$
are cubic (Kerr) nonlinear coefficients. In the absence of the Kerr
nonlinearities, these equations are the same as those use by Karamzin and
Sukhorukov in 1974 \cite{KS74} to obtain their famous $\chi^{(2)}$ soliton
solution, in which both the FH and SH fields are proportional to sech$^2$. A
detailed analysis of the higher-order soliton solutions in that model with
purely quadratic nonlinearity can be found in Ref. \cite{Drum}. However the
solutions that we will consider here are in a different class, in that the
FH field will be more like a sech than a sech$^2$.

The particular case of Eqs. (\ref{SHGPDE}) with $\delta =-1/2$ is specially
important, as it corresponds, with $t$ replaced by the transverse coordinate 
$x$, to a second-harmonic-generation model in the spatial domain (in fact,
in a nonlinear planar optical waveguide). In this special case, the model (%
\ref{SHGPDE}) is Galilean invariant, which allows one to generate a whole
family of ``moving'' solitons from the single zero-walkoff one \cite
{EtPeLeMa:95}. At all other values of $\delta $, construction of a
``moving'' (nonzero-walkoff) soliton is a nontrivial problem.

Stationary solutions to Eq. (\ref{SHGPDE}) are sought for in the form $%
u=U(t)\exp (ikz)$, $v=V(t)\exp (2ikz)$, where $k$ is real, and $U,V$ satisfy
ODEs 
\begin{equation}
\begin{array}{rcl}
(1/2)U^{\prime \prime }-kU+U^{\ast }V+\gamma _{1}(|U|^{2}+2|V|^{2})U & = & 0,
\\ 
-(1/2)\delta V^{\prime \prime }+(q-2k)V+(1/2)U^{2}+2\gamma
_{2}(|V|^{2}+2|U|^{2})V & = & 0.
\end{array}
\label{SHGODE}
\end{equation}
In Ref. \cite{YaMaKa:99}, ES solutions to these equations were found for $%
\delta >0$ and $\gamma _{1,2}<0$ (which implies anomalous and normal
dispersions, respectively, at FH and SH, and self-defocusing Kerr
nonlinearity. Alternatively, the same case may be physically realized as the
normal and anomalous dispersions at FH and SH and self-focusing Kerr
nonlinearity). In the same work, stability of these ESs was studied in
detail.

Here we shall present new results for $\delta <0$, which corresponds to a
more common case where the dispersion has the same sign at both harmonics.
The results can be naturally displayed in the form of curves of ESs in the $%
(\delta ,q)$ or $(q,k)$ parameter planes, see Figs. \ref{F:dm1} and \ref
{F:dm2} below.

\section{Existence}

\label{sec:2}

We now give three distinct explanations of why and how an ES may exist.

\subsection{Nonlinearizablity}

\label{sec:2.1}

\begin{figure}[tbp]
\epsfxsize 5in \centerline{\epsffile{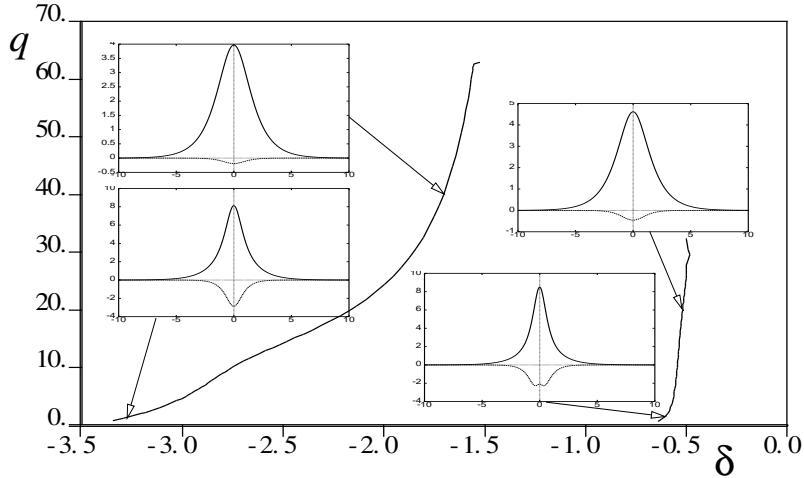}}
\caption{Two branches, in the $(\protect\delta,q)$ plane, of ES solutions to
the second-harmonic generation model (\ref{SHGPDE}) with $k=0.3$ and $%
\protect\gamma_1=\protect\gamma_2=0.05$. The insets depict the profiles of $%
U(t)$ (positive component) and $V(t)$ (negative one).}
\label{F:dm1}
\end{figure}

\begin{figure}[tbp]
\epsfxsize 4in \centerline{\epsffile{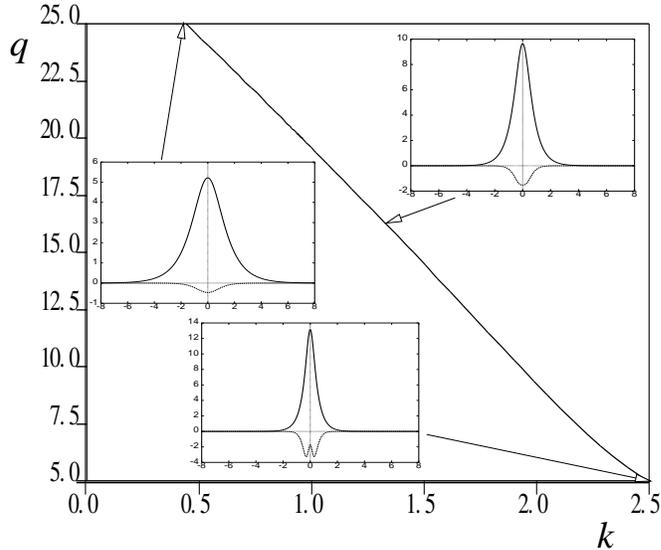}}
\caption{The branch of $\protect\delta=-0.5$ ESs in the $(k,q)$-plane
corresponding to the right-hand branch in Fig. \ref{F:dm1} (with $\protect%
\gamma_1=\protect\gamma_2=0.05$), the insets similarly showing the soliton's
profile.}
\label{F:dm2}
\end{figure}

Consider first the ODE system (\ref{SHGODE}). It is important to note that
the system gets fully decoupled in the linear approximation, the
linearization of its second equation immediately telling one that no true
soliton (with exponentially decaying tails) can exist inside the continuous
(radiation-mode) SH spectrum. However, it may happen that the tail of the
soliton's SH component decays at the same rate as the {\em square} of the
tail of the FH component. In that case, the second equation of the system (%
\ref{SHGODE}) is {\it nonlinearizable}, which opens the way for the
existence of truly localized solitons inside the continuous spectrum. Note
from the profiles of the solutions in Figs. \ref{F:dm1} and \ref{F:dm2},
that the $V$-component appears to decay to zero much faster than $U$, in
accordance with this nonlinearizability property.

Let us remark on some qualitative features of the ES branches presented in
Fig.\ \ref{F:dm1}. First, the computed branches appear to end in ``mid air".
At the low $q$ end, we have a boundary. The condition for the origin of the
ODE system (\ref{SHGODE}) to be a saddle-centre is $k<q/2$. For $q$ below
this limit, one may linearize (\ref{SHGODE}), and find regular
(non-embedded) solitons, for any $q$, provided $k>0$. These latter solitons
continuously match into the embedded solitons at $q=k/2$. Thus, since we
took $k=0.3$, we have that both branches of ESs will end at $q=0.6$. At the
high $q$ end, there were numerical difficulties in continuing the branches,
using the software AUTO \cite{AUTO}, for precisely the same reasons why the
computations presented in Fig. \ref{F:groves_gen} were difficult at $a=0^-$
for $b<0$. As one can see in the figure, one does have the amplitudes of
both $U$ and $V$ tending to zero (note also that the $V$-component is much
smaller, in accordance with the nonlinearizability principle).

Second, note that for high $q$, both ES solutions appear to be single-humped
and fundamental. But upon going to smaller $q$, where the amplitudes of both 
$U$ and $V$ become larger, it is apparent that the right-hand branch is a
higher-order state, with a superimposed ripple in the $V$ component (akin to
the second branch in Figs. \ref{F:groves_gen} and \ref{F:Mak0}).

Finally, we remark that only this second branch passes through the
physically significant value of $\delta=-1/2$. For the spatial ESs,
corresponding to $\delta =-1/2$, one is interested in how the solutions will
vary as a function of $k$ as well. That is shown in Fig. \ref{F:dm2}. Again,
the low-$q$ limit corresponds to the boundary of the saddle-centre region is 
$k=q/2$.

\subsection{Homoclinic orbits to saddle-centres}

\label{sec:2.2}

Next, let us try to understand why ESs should be of codimension-one. We
start from a general fourth-order Hamiltonian and reversible (invariant with
respect to reflections of time and `velocity' variables) ODE system.
Linearizing about the zero solution (the origin), we assume the system to
have a saddle-centre equilibrium. Thus, we have 
\begin{eqnarray}
\dot{x}=f(x), \quad x\in \Rset^{4}, \quad \exists R,&\quad& R^{2}=\mbox{id}
,\quad Rf(Rx)=-f(x),  \label{revsys} \\
\exists H(x)=\mbox{const. along solutions}, &\quad & \mbox{eigs}
(Df(0))=\{\lambda ,-\lambda ,i\omega ,-i\omega \}.  \nonumber
\end{eqnarray}
For example, for the ODE (\ref{grovesmodel}) the reversibility operator $R$
is given by 
\[
(u,u^{\prime},u^{\prime\prime},u^{\prime\prime\prime}) \mapsto
(u,-u^{\prime},u^{\prime\prime},-u^{\prime\prime\prime}),
\]
and for the system (\ref{SHGODE}) by 
\[
R: (U,U^{\prime},V,V^{\prime}) \mapsto (U,-U^{\prime},V,-V^{\prime}).
\]
A homoclinic orbit connecting such a saddle-centre equilibrium to itself is
formed by a trajectory simultaneously belonging to the one-dimensional
unstable and stable manifolds of the origin. Both of these manifolds lie in
a 3D phase space $H(0)$. Therefore, were it not for the reversibility, such
homoclinic orbits would be of codimension-two in general \cite{Le:91}, since
we require the coincidence of two lines in the three-dimensional space. But, 
{\em reversible} homoclinic solutions (i.e., solutions that somewhere
intersect the fixed-point set of the reversibility) are of codimension-one.
This is because the unstable manifold and $\mbox{fix}(R)\cap H(0)$ are both
one-dimensional, and we only require a point intersection between them.
Hence, varying two parameters, we should expect to see ESs occurring along
lines in the corresponding two-dimensional parameter plane. Moreover, the
solutions themselves must be reversible; asymmetric ESs would be of a higher
codimension still.

Mielke, Holmes and O'Reilly \cite{MiHoOR:92} proved a general theorem valid
in the neighborhood of such a curve in the parameter plane of reversible
saddle-centre homoclinic orbits: under a sign condition, essentially
governing how reversibility and the Hamiltonian interact, they showed that
there will be an accumulation of infinitely many curves of $N$-pulse ``bound
states" of the primary homoclinic orbit, for each $N>1$. Note that, for
systems that are reversible and also have odd symmetry (such as (\ref
{GTM4thODE})), the sign condition is always satisfied by virtue of the
system's admitting both reversibilities $R$ and $-R$ (i.e., the model (\ref
{GTM4thODE}) is symmetric too under $U\to -U$, and hence is also invariant
under the reversibilities $(U,U^{\prime}) \to (U^*,-{U^{\prime}}^*)$ {\em and} 
$(U,U^{\prime}) \to (-U^*,U^{\prime})$). Here, the sign condition determines
whether there are ``up-up" or ``up-down" bound states. In Ref.\ \cite
{ChMaFr:98}, a large number of the bound states of the ``up-down" type were
found for the generalized massive Thirring model (\ref{GTM4thODE}) in
agreement with this theory. We also remark that Buryak and co-workers (see 
\cite{Bu:95}) found a similar discrete sequence of bound states of
`nonexistent' dark solitons in a SHG model and a higher-order NLS equation,
however none of these solutions were linearly stable.

So far, in every case that we have investigated, we have found the
higher-order ESs to be unstable against linear perturbations. A distinction
should be stressed between what one would call ``bound-states" --- which are
like several copies of a fundamental soliton placed end to end --- and the
higher-order solitons, such as those displayed in Figs. \ref{F:groves_gen}
and \ref{F:Mak0}, which are like a fundamental with internal ripples. At the
moment, there seems to be no connection between these states, although it is
still may happen that, as some parameter is varied, a continuous branch may
connect solutions of the two different types.

\subsection{A singular limit}

\label{sec:2.3}

We now look at a mechanism which explains how fundamental ESs (possibly with
ripples) may appear from the singular limit $\lambda \to 0^+$. Such a limit
for the general class of systems (\ref{revsys}) has been studied using the
normal-form theory by Lombardi \cite{Lo:96,Lo:97b,Lo:98}, incorporating
careful estimation of various exponentially small terms (cf.\ related
results obtained using exponential asymptotics, e.g. \cite{GrJo:95,GrCo:96}). 
A crucial additional ingredient we shall add to the Lombardi's work is
that $\lambda $ and $\omega$ are assumed to play the role of two independent
parameters. We provide here only an oversimplified sketch, more details will
appear elsewhere \cite{Ch:99}.

The appropriate normal form is \cite{Lo:96} 
\begin{equation}
\begin{array}{rcll}
\dot{x}_{1} & = & x_{2}, &  \\ 
\dot{x}_{2} & = & \lambda ^{2}x_{1}-(3/2)x_{1}^{2}-b_1(x_{3}^{2}+x_{4}^{2})
& + \rho N_{2}(x;\lambda ), \\ 
\dot{x}_{3} & = & -x_{4}(\omega + b_2 x_{1}) & + \rho N_{3}(x;\lambda ), \\ 
\dot{x}_{4} & = & x_{3}(\omega + b_2 x_{1}) & + \rho N_{4}(x;\lambda )
\end{array}
\label{8.norm_fullsys}
\end{equation}
where $b_{1,2}$ are $\omega$-dependent constants to be determined for a
particular system, and $N_i$ are higher-order (remainder) terms that break
(for nonzero $\rho$) the completely-integrable structure of the truncated
normal form. It is not difficult to see that the truncated system possess a $%
\mbox{sech}^2$-like homoclinic connection to the origin, whose amplitude is $%
O(\lambda)$. The key question is whether this homoclinic orbit persists
under the inclusion of the remainder terms. This question can be posed in
terms of the vanishing of a certain Melnikov integral (whose vanishing
measures the splitting distance between the stable and unstable manifolds),
which, after a lengthy calculation \cite{Lo:97b}, can be written as 
\[
I=\frac{\rho }{\lambda ^{2}}\exp (-\omega \pi /\lambda )(\Lambda
(N_{3},N_{4},\omega )(1+O(\lambda)+O(\rho)) , 
\]
where $\Lambda (N_{3},N_{4},\omega )$ can be computed explicitly for each
monomial which is pure in $x_{1}$ and $x_{2}$ in the Taylor-series expansion
of either $N_{3}$ or $N_{4}$.

Now, something beautiful happens because, for each such monomial, $\Lambda$
turns out to be \cite{Lo:97b} a (single-signed $\omega $-dependent) constant
multiple of either a Bessel function, $\Lambda \sim J_{n}(4\sqrt{\omega b_2})
$ for $b_2>0$, or modified Bessel function $\Lambda \sim I_{n}(4\sqrt{\omega
|b_2|})$ for $b_2<0$, for some integer $n$. Recall the basic properties of
the Bessel functions, according to which $J_{n}(x)$ has infinitely many
zeros for $x>0$, whereas the modified Bessel function $%
I_{n}(x)=i^{-n}J_{n}(ix)$ has no zeros. Hence, notice the crucial role
played by the coefficient $b_2$ (in the truncated, scaled normal form (\ref
{8.norm_fullsys})) in the case when the remainder $N$ consists of a single
monomial in $(x_{1},x_{2})$. If $b_2>0,$ there will be an infinite number of 
$\omega$ values corresponding to zeros of $\Lambda $ and hence homoclinic
solutions will exist for small $\rho $ and $\lambda$. However, if $b_2<0,$
then $\Lambda $ is strictly non-zero and hence there are no homoclinic
solutions to the origin.

The coefficient $b_2$ is easy to calculate in the examples with the
quadratic nonlinearities, such as the extended 5th-order KdV model (\ref
{grovesmodel}) with $\mu =1$ (see \cite{Ch:99} for details). There it is
found $b_2>0$ and this entirely explains the approximately periodic sequence
of points on the negative $b$-axis at which the homoclinic solutions
bifurcate from $a=0$ with the zero amplitude, as shown in Fig. \ref
{F:groves_gen}. In contrast, for $\mu =-1$, $b_2$ is negative and no ES
bifurcates from $a=0$. For models with purely cubic nonlinearity, such as
the generalized Thirring model (\ref{GTM4thODE}), it can be shown that all
quadratic coefficients in the normal form (\ref{8.norm_fullsys}) vanish,
hence a new, odd-symmetric normal form would need to be studied. This is
left for future work. We mention also that for the SHG model (\ref{SHGODE}),
although there are quadratic terms in it, the coefficient $b_2$ is
identically zero, so this analytic technique gives no information.

\section{Stability}

\label{sec:3}

\begin{figure}[tbp]
\epsfxsize 5in \centerline{\epsffile{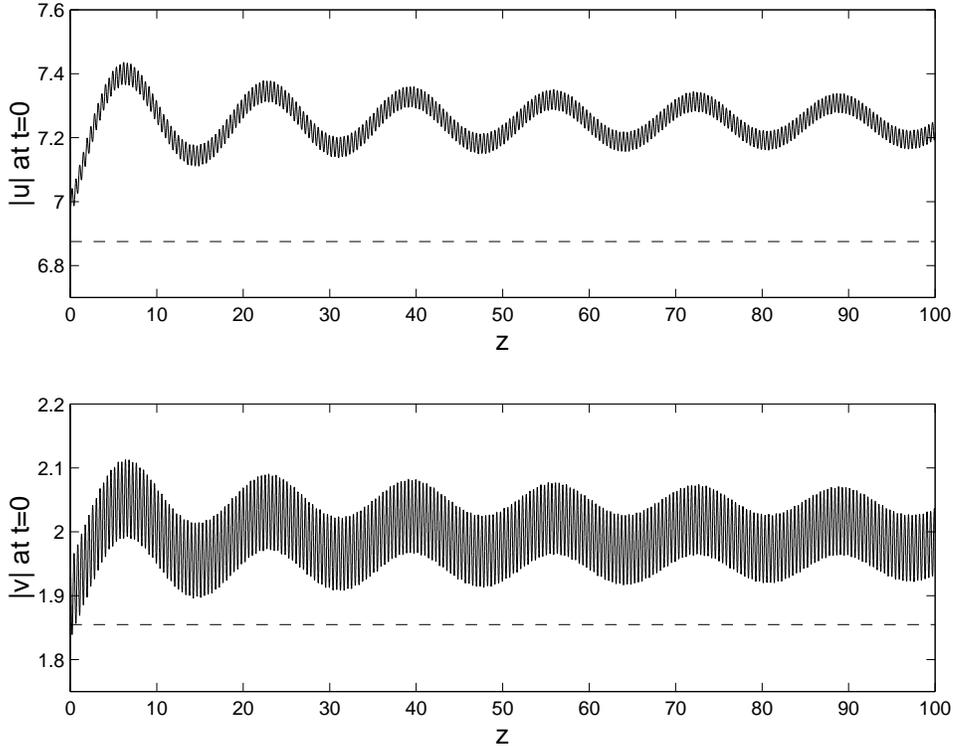}}
\caption{Evolution of a perturbed soliton with positive energy perturbation (%
$\protect\alpha_1=0.1, \protect\alpha_2=-0.1$ in (\ref{ici})) for (\ref
{SHGPDE}) from the left-hand fundamental branch of Fig.\ \ref{F:dm1}}
\label{F:jianke1}
\end{figure}

The stability of the embedded solitons in the SHG model has been studied
both numerically and analytically in \cite{YaMaKa:99}. It was shown that the
fundamental (single-hump) ES is linearly stable, but nonlinearly
semi-unstable, while all the multi-humped ES are linearly unstable. In the
semi-stability analysis of the fundamental ES, a crucial role is played by
the energy, as ESs are {\em isolated} solutions with uniquely determined
values of the energy, with the adjacent delocalized soliton states, on
either side, having an infinite energy. It was argued in \cite{YaMaKa:99}
(see also a similar argument given by Buryak \cite{Bu:95}) that
perturbations which slightly increases the fundamental ES's energy can be
safe, while perturbation which decreases the energy inevitably triggers a
slow (sub-exponential) decay of the soliton into radiation. Thus, the weak
instability of an ES is {\em one-sided}. This fact also follows from a
simple argument that the usual exponential instability is always dual-sided.
Equivalently, the usual instability is linear, while the weak one-sided
instability of ESs must be {\em nonlinear}.

\begin{figure}[tbp]
\epsfxsize 4in \centerline{\epsffile{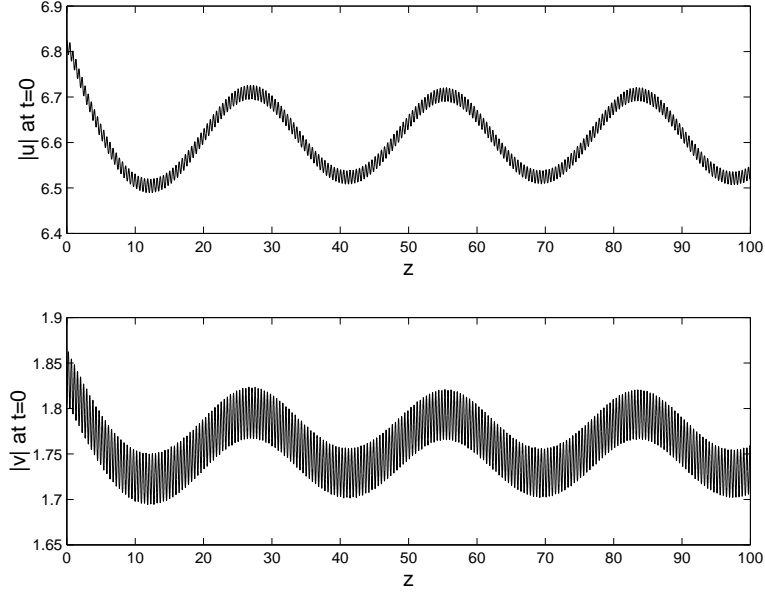}}
\caption{Evolution of a perturbed soliton with a negative energy
perturbation ($\protect\alpha_1=-0.05, \protect\alpha_2=0.05$ in Eq. (\ref
{ici})) in the model (\ref{SHGPDE}). The soliton belongs to the left-hand
fundamental branch in Fig. \ref{F:dm1}}
\label{F:jianke2}
\end{figure}

The situation is the same in the present case where $\delta < 0$. A study of
the linearized equation around the ES shows that the fundamental ES branch
(the left one in Fig. 3) is linearly stable, while the branch of
multi-humped ES (the right one in Fig. 3) is linearly unstable. The
semi-stability argument for the fundamental ES branch also applies here.
Positive energy perturbations can be safe, while negative energy
perturbations trigger decay of ES. However, as we shall see below, for $%
\delta < 0$, this decay seems to be significantly slower than that found in
Ref. \cite{YaMaKa:99} for $\delta > 0$. As was done there, we numerically
simulated the system $(\ref{SHGPDE})$, with the initial data 
\begin{equation}
u(0, t)=U(t)+ \alpha_1 \mbox{sech} 2t, \hspace{0.5cm} v(0, t)=V(t)+ \alpha_2 %
\mbox{sech} 2t\,,  \label{ici}
\end{equation}
where $U(t)$ and $V(t)$ is an ES solution on the left-hand branch of Fig. 
\ref{F:dm1} (the values are $\delta=-2.9292$, $q=6.0556$, $k=0.2936251$ and $%
\alpha_1=\alpha_2=0.05$). Figs. \ref{F:jianke1} and \ref{F:jianke2} depict,
respectively, the effects of the positive-energy and negative-energy
perturbations. In both cases, we observe fast oscillations on top of slow
ones in the evolution of $|u|$ and $|v|$ at $t=0$. These oscillations are
very similar to those reported in \cite{EtPeLeMa:95} for perturbed solitons
in the standard SHG system with no cubic nonlinearity (note that those
solitons are ordinary ones, rather than ES). In that case, the fast
oscillations were attributed to an intrinsic mode, while the slow
oscillations were attributed to beatings between the intrinsic mode and a 
{\it quasi-mode}. The latter one is localized in the FH component, but
resonates with the continuous spectrum in the SH component. Both
oscillations could last for a very long time, even though they were expected
to eventually decay due to a very weak radiation damping.

We believe that similar mechanisms are also at work in our model. However,
there are important differences because of the fact that the solitons in our
model are embedded, and those in the model considered in Ref. \cite
{EtPeLeMa:95} were not. On the other hand, there are important similarities
because perturbations of an ES naturally resonate with the continuous
spectrum of the SH component, and we do see such slow oscillations as well.
Although according to the semi-stability argument, a negative perturbation
of an ES, as shown in Fig. \ref{F:jianke2} would eventually decay, it is a
very slow decay. Detailed examination of the numerical solutions shows that
the central pulse (the $v$ component) in Fig. \ref{F:jianke2} keeps shedding
oscillating tails into the far field. However, the tail amplitudes are
extremely small (about 0.001 or smaller). This is why the expected decay of
the perturbed ES is not obvious in that figure. Because of this, the
actually observed evolution is dominated by the beating and internal
oscillations, just as in Fig. \ref{F:jianke1} with a positive perturbation,
and as in Ref. \cite{EtPeLeMa:95}. It will take an extremely long time for
the pulse in Fig. \ref{F:jianke2} to show considerable decay. In fact, it is
clear that the ESs in the present model with $\delta < 0$ are {\it virtually
stable}, when compared to the previously considered case \cite{YaMaKa:99}, $%
\delta >0$, where the semi-instability was a really observed feature. The
relative stability of an ES for $\delta < 0$ is a new result reported in the
present paper, and its importance is quite obvious.

\begin{figure}[tbp]
\epsfxsize 5.5in \centerline{\epsffile{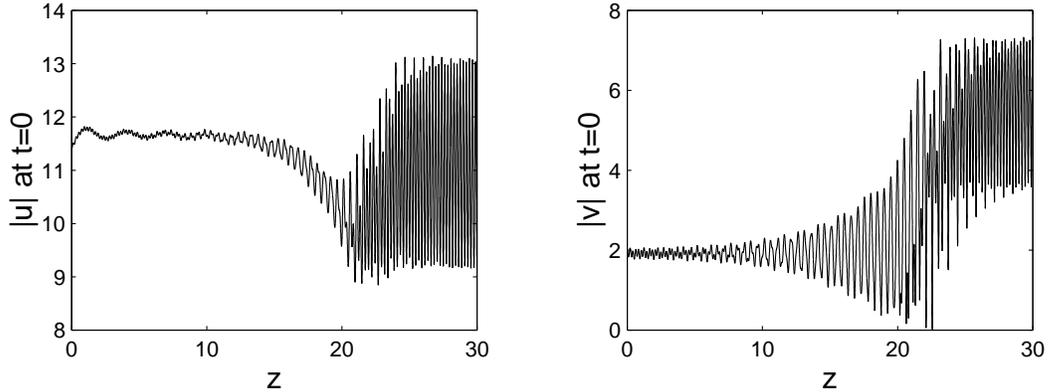}}
\caption{The linear instability of a perturbed soliton belonging to the
right-hand branch in Fig. \ref{F:dm1} at $\protect\delta=-0.5$, $\protect%
\gamma_1=\protect\gamma_2=0.05$, $q=10$ and $k=1.9298$}
\label{F:jianke3}
\end{figure}

If the ES were linearly unstable, the semi-stability argument would not
apply. This is the case for the solutions investigated in \cite{Bu:95}, and
also for the right-hand branch depicted in Fig.\ \ref{F:dm1}, which
corresponds to the case of a nonlinear planar optical waveguide in the
spatial domain. To verify this, we have performed a time integration of the
PDE, the results of which are presented in Fig. \ref{F:jianke3}. One can see
the onset of a violent exponential instability. We have verified that this
solution does have exponentially unstable eigenmodes, from a numerical study
of the linearization of Eq. (\ref{SHGPDE}) about this solution.

In view of these results we conjecture the following. Fundamental ESs are,
in general, linearly neutrally stable but semi-stable nonlinearly. The
higher-order ESs (which usually have internal ripples in their profiles)
are, generally, linearly unstable. Preliminary results for the extended
5th-order KdV (\ref{5thKdV}) indicate qualitatively the same properties.
This would also accord with previous numerical results for higher-order NLS
equations that an isolated fundamental ES \cite{FuEs:96} is semi-stable
whereas multihumped bound-states \cite{Bu:95} are exponentially unstable.

To conclude this section, we notice that multihumped solitons (which are 
{\em not} a subject of the present paper) of the ordinary (nonembedded) type
were found in many models, see, e.g., early works \cite{Se80,Se81} dealing
with the Langmuir waves in plasmas, and a recent work on the resonant
three-wave interactions

\section{Moving embedded solitons}

\label{sec:4}

\begin{figure}[tbp]
\epsfysize 7cm \centerline{\epsffile{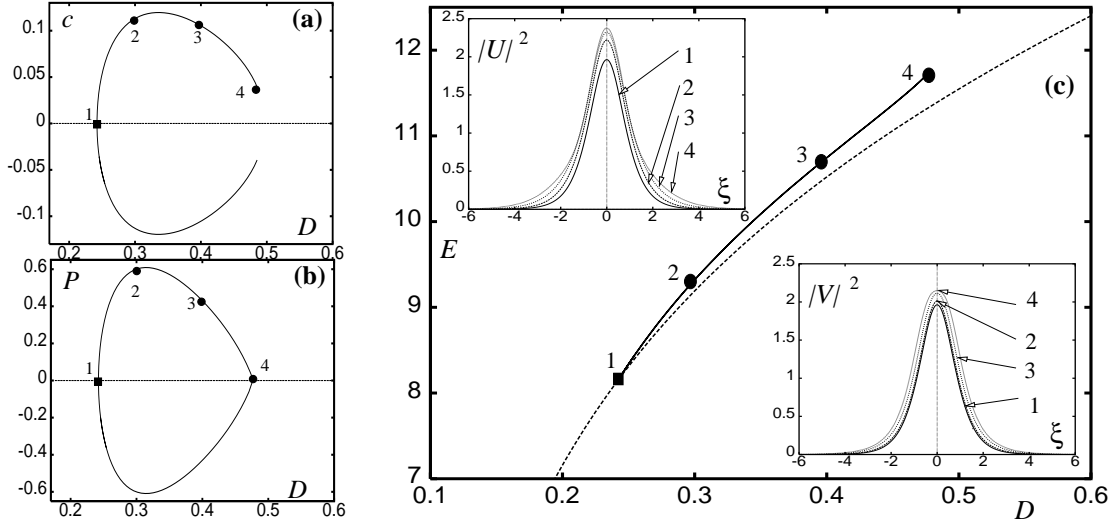}}
\caption{{\protect\small A branch of moving ESs bifurcating from one of the
three simplest branches of the fundamental quiescent ESs in Eqs. (\ref{GTM})
with $\protect\sigma=1/2$: the velocity (a), momentum (b), and energy (c)
vs. $D$, with insets showing the shape of the moving solitons at labeled
points.} }
\label{F:dij1}
\end{figure}

We will now discuss a possibility that ESs in the generalized Massive
Thirring model (\ref{GTM}) may be moving at a non-zero velocity $c$ (this
discussion applies also to ESs in the three-wave system (\ref{MakPDE}) with
a non-zero walkoff). In this case, the reduction of the 8th-order ODE to a
4th-order one is no longer possible. Moreover, since the 8th-order system is
obtained by separating the real and imaginary parts of a 4th-order complex
system, the spectrum will have a double degeneracy when $c=0$. This means
that, in the parameter region of interest in the $(\chi ,c,D)$-space, the
linearization yields four pure imaginary eigenvalues, plus two with positive
real parts and two with negative ones. A similar counting argument, as in
Section (\ref{sec:2.2}), shows that reversible homoclinic orbits to such
equilibria are of {\em codimension two}. Hence ESs lie on one-dimensional
curves in the three-parameter space. Alternatively, this property can be
explained as follows: in addition to the energy $E$, the full system also
preserves the momentum $P$, so we can view a moving ES as being isolated in 
{\em both} invariants, i.e., the ES solution family is described by curves $%
E(D)$ and $P(D)$. Finally, we find that such curves can be found naturally
as bifurcation points at $c=0$ from curves of the zero-velocity ESs.

\begin{figure}[tbp]
\epsfxsize 6in \centerline{\epsffile{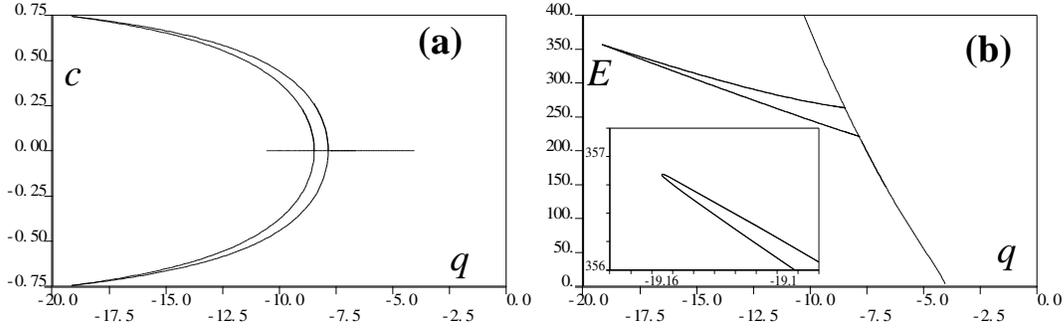}}
\caption{Two branches of ``walking" ($c\neq 0$) embedded solitons
bifurcating from the ground-state $c=0$ branch (solid horizonal line) for
the three-wave model $(\ref{MakPDE})$: (a) the walkoff $c$, and (b) the
energy flux $E$ vs. the mismatch $q$. The inset in (b) shows that the two
branches meet on the left and merge in a typical tangent bifurcation.}
\label{F:mak4}
\end{figure}

Fig. \ref{F:dij1} shows one such bifurcation occurring from one of the three
branches of fundamental ESs for the Thirring model. The ES branches
terminate (as at point ``4") and beyond that, ES become ordinary
(non-embedded) solitons (i.e., where the saddle-centre equilibrium becomes a
pure saddle with only real eigenvalues).

Similarly, Fig. \ref{F:mak4} shows two branches of moving ESs bifurcating
from the ground-state ES (the nearly vertical line) in the three-wave model.
Note that the two bifurcating curves become globally connected on the left,
through a regular tangent (fold) bifurcation.

Finally, let us turn to the second-harmonic model (\ref{SHGPDE}). As it was
mentioned above, precisely at the value $\delta = -1/2$, this model gives
rise to moving solitons in a trivial way, via the Galilean transformation.
Search for moving ES in this model with $\delta \neq -1/2$ is the subject of
ongoing work.

\section{Conclusion}

\label{sec:5}

In this paper, we have given a brief overview of recent results that
establish the existence of isolated (codimension-one) solitons embedded into
the continuous spectrum of radiation modes. A necessary condition for the
existence of the embedded solitons is the presence of (at least) two
different branches in the spectrum of the corresponding linearized system,
so that one branch can correspond to purely imaginary eigenvalues, and
another to purely real ones. The fundamental (single-humped) embedded
solitons are always stable in the linear approximation, being subject to a
weak sub-exponential one-sided instability. Moving embedded solitons may
also exist as codimension-two solutions. Moreover, bound states in the form
of multi-humped embedded solitons exist too, but they are linearly unstable.

We should remark that the work presented and reviewed 
here does not necessarily
represent the very first time that the existence of ESs have been
established in any physical model. For example, as stated earlier,
multi-humped dark ESs were already observed in a SHG model and
for higher-order NLS equations \cite{Bu:95} where they were found to
be unstable. Fundamental ESs were also found for the latter with competing
cubic and quintic nonlinearities \cite{FuEs:96} and their existence
was also suggested for a coupled KdV
system \cite{GrCo:96}, without evidence to suggest their stability.
Also recent work \cite{BoKo:94,ChKi:00}
has shown the existence of front solutions to higher-order
Frekel-Kontorova models which are in resonance with the linear
spectrum; so called `embedded kinks'. Nevertheless, we believe the results
presented in this paper --- the new stability results for fundamental
ESs in Section 4, together with the three distinct views given in
Section 3 as to why ESs should exist, and the similarities found in
Section 2 between their existence properties for four distinct models
--- provide a new theory for ESs as a phenomenon in their own right.

Moreover, the very fact that ESs are isolated states suggest that they
may find potential application in photonics, such as in
all-optical switching. For example, taking the examples in Sections
2.2 and 2.3, switching from one ES state to a neighboring one with a
smaller energy might be easily initiated by a small perturbation, in
view of the semi-stability inherent to ESs. Switching between two
branches of moving ESs with $c\neq 0$ might be quite easy to realize
too, due to the small energy-flux and walkoff differences between
them. There remains much work to be done in investigating these
potential aplications further.

Many issues concerning the embedded solitons remain open and are a subject
of ongoing investigations. In particular, immediate questions arise
concerning moving ESs, and interactions caused by collisions between them.

We appreciate a useful discussion with A. Buryak. The research of JY was
supported in part by the NSF and the AFOSR. The research of DJK was
supported in part by the AFOSR. A collaboration between ARC and BAM was
supported by a fellowship granted by the Benjamin Meaker Foundation through
the University of Bristol. ARC is suppoted by an Advanced Fellowship from
the EPSRC.

\bibliographystyle{unsrt}
\bibliography{embed}

\end{document}